\documentclass[11pt]{article}
\usepackage{moriond,epsfig}

\def\be{\begin{equation}}
\def\ee{\end{equation}}
\def\bea{\begin{eqnarray}}
\def\eea{\end{eqnarray}}

\begin{document}
\vspace*{4cm}
\title{TOWARDS AN ULTRA-STABLE OPTICAL SAPPHIRE CAVITY SYSTEM FOR TESTING LORENTZ INVARIANCE}

\author{M. NAGEL and A. PETERS}

\address{Humboldt-Universit\"{a}t zu Berlin, Institut f\"{u}r Physik, Newtonstr. 15,\\
12489 Berlin, Germany}

\maketitle\abstracts{
We present a design for an ultra-stable cryogenically cooled sapphire optical cavity system, with fractional frequency stability better than  $1\times10^{-16}$ at one second integration time. We plan to use such ultra-stable cavities to perform a test of the isotropy of light propagation at the $10^{-20}$ level.}

\section{Motivation}
Many experimental and technical applications, e.g. optical atomic clocks, demand ultra-stable cavity systems for laser frequency stabilization. Nowadays, the main limiting factor in frequency stability for room temperature resonators has been identified to be the displacement noise within the resonator substrates and mirror coatings due to thermal noise. Different approaches are being proposed right now to lower the influence of thermal noise, such as using higher order modes, longer cavities, or new types of coating materials. A straightforward method is to cool down the resonators to cryogenic temperatures. Following this approach, we will set up an advanced cryogenic optical resonator system using specially designed sapphire cavities with the goal to reach a relative frequency stability of better than $1\times10^{-16}$ up to long integration times.

\section{Next Generation Cryogenic Optical Cavity System}

\begin{figure}
\epsfig{figure=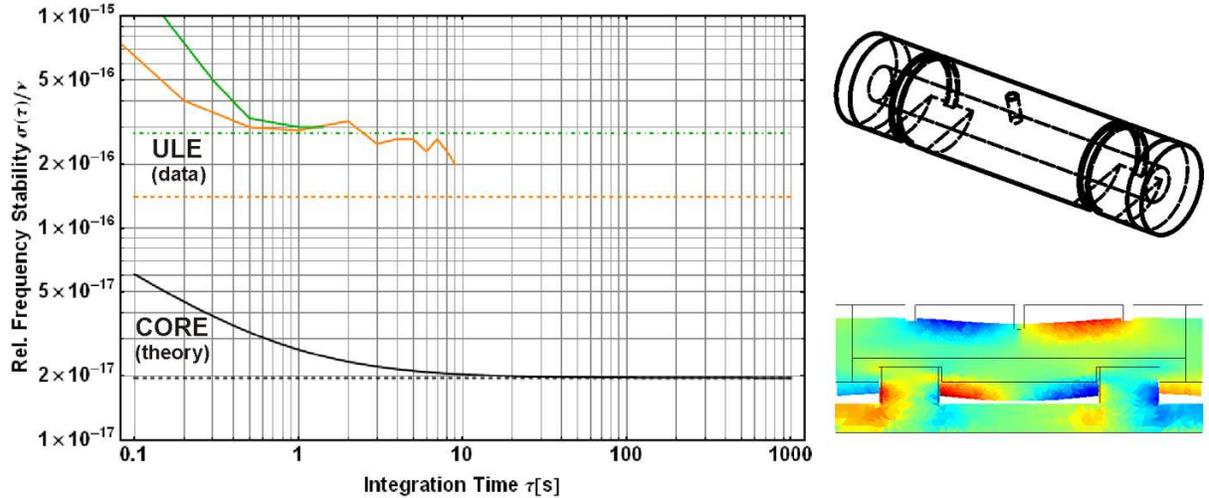,width=\textwidth}
\caption{Left: Comparison of measured and predicted rel. freq. stability of the best room temperature (ULE, Young et. al [NIST 1999] green line and Jiang et. al [NIST 2010] orange line) and proposed cryogenic optical resonators (CORE, black line). The dashed lines show the theoretical thermal noise limit. Upper Right: Sketch of an optimized design for a CORE. Lower Right: FEM simulation of vertical vibration (gravitational) induced bending (deformation scaled up by a factor of $10^{10}$).}
\label{fig:newCORE}
\end{figure}

Two normally opposing requirements need to be matched in designing a cryogenic resonator and its mounting structure: high thermal conductivity towards the liquid helium bath and low mechanical coupling of the optical path length to vibrations. We used FEM computations to optimize the design of a resonator made of sapphire which reduces the influence of vertical and horizontal vibrations and at the same time features large thermal contact areas for the mounting structure (see Figure \ref{fig:newCORE}).

Calculations on the shot noise and thermal noise level show that the theoretical frequency stability at a temperature of 4.2K is in total an order of magnitude better than the best ever value obtained with an optical resonator system (see Figure \ref{fig:newCORE}). We will also implement novel measures to enhance the long term performance of the optical cavity system in order to maintain the potential relative frequency stability below $1\times10^{-16}$ up to long integration times.

As a future perspective, we plan to exchange the Ta$_{2}$O$_{5}$/SiO$_{2}$ mirror coatings, which are the thermal noise limiting source for these cryogenic optical sapphire resonators, with monocrystalline coatings composed of Al$_{x}$Ga$_{1-x}$As.\cite{cole} Those crystalline coatings will further reduce the stability limiting effects of thermal noise by more than an additional order of magnitude.

\section{Testing Lorentz Invariance}

We plan to use the ultra-stable cavities to perform a laboratory-based test of Lorentz invariance - a basic principle of the theories of special and general relativity. While both theories developed by Einstein play an integral part in modern physics and in today's ordinary life, there have been claims that a violation of Lorentz invariance might arise within a yet to be formulated theory of quantum gravity.

The cavities will be arranged in a Michelson-Morley configuration and continuously rotated with a rotation period between 10s and 100s for more than one year using a custom-made high-precision low noise turntable system made of granite. The sensitivity of this setup to violations of Lorentz invariance should be in the $10^{-19}$ to $10^{-20}$ regime. This corresponds to more than a 100-fold improvement in precision of modern Michelson-Morley type experiments.\cite{sh}

Furthermore, ultra-stable cryogenic microwave whispering gallery resonators will be added to the experiment in collaboration with the University of Western Australia. With this co-rotating microwave and optical resonator setup we will be able to search for additional types of Lorentz violating signals.

\end{document}